\documentclass[12pt]{elsart}

\usepackage{graphicx}  
\usepackage{subfigure}  
\usepackage{epstopdf} 
\newcommand{\rme}{{\rm e}}
\newcommand{\rmd}{{\rm d}}
\newcommand{\ds}{\displaystyle}

\usepackage{amssymb}
\usepackage{times}

\begin{document}

\begin{frontmatter}



\title{Optical properties of  PZT thin films deposited on a ZnO buffer layer.}


\author{T. Schneider, D. Leduc, J. Cardin, C. Lupi, N. Barreau$^\star$ and H. Gundel}

\address{Universit\'e de Nantes, Nantes Atlantique Universit\'es, IREENA, EA1770, 
Facult\'e des Sciences et des Techniques, 2 rue de la Houssini\`ere - BP 9208, Nantes, F-44000 France.}
\address{$^\star$Universit\'e de Nantes, Nantes Atlantique Universit\'es, LAMP, EA3825, 
Facult\'e des Sciences et des Techniques, 2 rue de la Houssini\`ere - BP 9208, Nantes, F-44000 France.}
\ead{dominique.leduc@univ-nantes.fr}

\begin{abstract}
 The optical properties of lead zirconate titanate (PZT) thin films deposited 
 on ZnO were studied by m-lines spectroscopy. In order to retrieve the refractive 
 index and the thickness of both layers from the m-lines spectra, we develop a numerical
  algorithm for the case of a two-layer system and show its robustness in the presence 
  of noise. The sensitivity of the algorithm of the two-layer model allows us to relate 
  the observed changes in the PZT refractive index to the PZT structural change due to 
  the ZnO interface of the PZT/ZnO optical waveguide.
\end{abstract}

\end{frontmatter}

\maketitle



\section{Introduction}

PZT  ferroelectric thin films exhibit interesting optical properties, such as 
large electro-optic effects (Pockels, Kerr), a high refractive index, and a high 
transparency at visible and infra red wavelength. These properties are promising for the realization 
of different applications in the field of integrated optics as sensors or components for optical
 communications (optical shutter,  waveguide, filter).
In order to obtain active devices profiting from the electro-optic properties, 
electrodes have to be integrated.
Most commonly they are  made of transparent conductive materials like doped ZnO or 
 indium tin oxide (ITO) when the absorption 
 of light is a critical factor. The present paper  studies 
  a bilayer made of PZT and a transparent  ZnO bottom electrode.
  It is generally known, that the crystallization behavior of ferroelectric thin films strongly 
  depends on the structural properties of the substrate and hence may be influenced by the existence
   of an interface layer. This has been shown particularly for the case of PZT thin films 
   elaborated on metal substrates using different conducting oxide interface layers \cite{seveno2001}.
    In the case of  the PZT/ZnO waveguide, we are interested 
   in the relation between the PZT  structural change and the optical properties of the films. 
   For this purpose we developed a characterization method for a two layer 
   planar waveguide based on prism coupler spectroscopy (or m-lines). M-lines spectroscopy~\cite{tien1970,ulrich1970,ulrich1973} is widely used 
in order to determine the refractive index and
the thickness of single layer homogeneous films.
The refractive index profile of inhomogeneous waveguide can also be reconstructed with m-lines spectroscopy
 by using
methods based on WKB approximation~\cite{white1976,chiang1985}. These methods are well suited in the case of
guides where the index profile can be described by an a priori known continuous function.
The case of multilayer guides where the  index profile contains abrupt discontinuities,
however, is less investigated.
Dispersion equations of guided modes for the two layer guides were initially introduced 
by Tien~\cite{tien1973} and several experiments 
were performed~\cite{stutius1977,matyas1991,aarnio1995,augusciuk2004}.
The reconstruction of the film parameters in these measurements were made with the help of
 an optimization method, the simplex algorithm in general, that minimizes
the differences between the measured effective indices and the calculated ones from
the dispersion equations. 
In our approach, we choose to   directly solve
the system of dispersion equations.
This requires to determine the roots of a system of  non linear equations with
five  unknowns (the refractive index and the thickness of the two  layers and the
order of the first mode in the spectrum). 
The first section of the present paper  is devoted to the study of the efficiency of this method.
We will especially examine the influence of the noise  in order to estimate the accuracy
of the results. Then in the second section, we will apply the method in order to 
firstly characterize   the ZnO  layer and  secondly  the two-layer waveguides.

\section{Characterization of two layers films using m-lines spectroscopy}

\subsection{Two layer dispersion equations}

\begin{figure}[htbp]
\centerline{
\includegraphics[width=5cm]{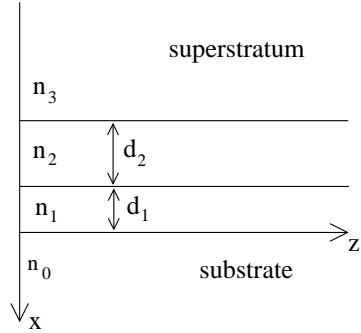}}
\caption{Two layer waveguide.}
\label{fig2}
\end{figure}

The analysis of m-lines spectra requires to know the dispersion
equation for the studied guide. Because of its compactness, the 
transfert matrix method~\cite{chilwell1984} is well suited 
 for analysing multilayer waveguides.  The geometry of the guide
is shown in the  figure~\ref{fig2}. The layer $j$ has a refractive
index $n_j$ and a thickness $d_j$. The layer $0$ corresponds to
the substrate and the layer $3$ to the superstratum. In our case,
$n_0=1.5169$ and $n_3=1$ at 632.8~nm wavelength. For TE modes, the electrical
field in the layer $j$ has only one  component along the $(Oy)$ axe~:
 $
 E_{jy}(x,z)=A_j\ \rme^{i\gamma x}\ \rme^{i\beta_m z}+ B_j \ \rme^{-i\gamma x}\ \rme^{i\beta_m z}
$
and the tangential component of the magnetic field is
$H_{jz}=i(\omega \mu_0)^{-1}\ \rmd E_{jy}/\rmd x$. In these expressions, $\omega$ is the angular frequency and $\beta_m$ is the propagation
constant of the m$^{\rm th}$ guided mode. It is usually written  as $\beta_m=kN_m$, where $k$
is the wavevector modulus in vacuum and $N_m$ the effective index.
The x component of the wavevector, $\gamma_j$, gives the nature of the waves in the layer $j$~:
$\gamma_{j}=k(\omega\mu_0)^{-1}\ |n_j^2-N^2|^{1/2}$ for travelling waves and $\gamma_{j}=i k(\omega\mu_0)^{-1}\ |n_j^2-N^2|^{1/2}$
for evanescent waves. In the following, we will call $a_j=k\ |n_j^2-N^2|^{1/2}$.

A transfert matrix $M_j$ is associated to each layer~:
\begin{equation}
\label{eq2}
M_j=
\left(\begin{array}{cc}
\cos (\omega\mu_0\gamma_j d_j) & \displaystyle \frac{i}{\gamma_j}\sin (\omega\mu_0\gamma_j d_j)\\
i\ \gamma_j \sin (\omega\mu_0\gamma_j d_j )& \cos (\omega\mu_0\gamma_j d_j)\\
\end{array}\right)
\end{equation}
The tangential component of the electric and the magnetic fields  $E_y$ and  $H_z$ must be continuous 
at the interface to satisfy boundary conditions. These conditions 
and the condition for obtaining guiding lead to the equation~: 
\begin{equation}
\label{eq3}
\left(\begin{array}{c} 1\\-\gamma_3 \end{array}\right) E_{3y}
      =M_2M_1 \left(\begin{array}{c} 1\\\gamma_0 \end{array}\right) E_{0y}
      =M\left(\begin{array}{c} 1\\\gamma_0 \end{array}\right) E_{0y}
\end{equation}
which has solutions only for~:
\begin{equation}
\label{eq4}
\gamma_3 m_{11}+\gamma_3\gamma_0 m_{12} + m_{21} + \gamma_0 m_{22}=0
\end{equation}
where $m_{ij}$ are the components of the matrix $M$.


 The refractive index of  PZT is known to be higher than that of  ZnO
 ($n_2>n_1$). Two kinds of guided waves are  possible in this case~:
 \begin{itemize}
 \item
 guided waves in layer 2,  evanescent waves  in the
 other layers
\begin{normalsize}
\begin{equation}
\label{eq5}
\mbox{\strut  \hspace*{-20mm}}
{a_{2}}\,{d_{2}}  - \;\mathrm{arctan} \left(   \frac{a_{3}}{a_{2}}   \right)
 - \;\mathrm{arctan} \left[{\ds \frac {a_1^2\,\mathrm{tanh}({a_{1}}\,{d_{1}}) + 
 a_{0} a_1 }{a_1 a_2 + { a_{0} a_2\, \mathrm{tanh}({a_{1}}\,{d_{1}})} }}    \right]
   -  \;m \,\pi =0
\end{equation}
\end{normalsize}
 \item
 guided waves in  layer 1 and layer 2,  evanescent waves  in the
substrate and in the superstratum
 \begin{normalsize}
 \begin{equation}
\label{eq6}
\mbox{\strut  \hspace*{-20mm}}
{a_{2}}\,{d_{2}}  - \;\mathrm{arctan} \left( \ds \frac{a_{3}}{a_{2}}   \right) + \;\mathrm{arctan} \left\{  \! {\ds 
\frac {{a_{1}}}{{a_{2}}}}\ {\mathrm{tan}\left[{a_{1}}\,{d_{1}} - \mathrm{arctan}\left(
{\ds \frac {{a_{0}}}{{a_{1}}}} \right)\right]}  \! 
 \right\}   -  \;m \,\pi =0
 \end{equation}
 \end{normalsize}
\end{itemize}
Equations \ref{eq5} and \ref{eq6} are the dispersion equations for the two layer
waveguide.

\subsection{Data analysis}

The characterization of a two layer waveguide requires to 
determine 5 unknowns~: the refractive indices and thicknesses of both layers
and the order ($m_1$) of the first mode appearing in the m-lines spectrum. Very often, the latter
parameter is  neglected and  the first mode in the spectrum is assumed
to be the fundamental mode. The low order modes, however,  are more difficult to excite
since they require an high incidence angle and  often 
spectra without the fundamental mode are observed. Therefore it seems indicated  not to make any
assumption on the order of the first mode and to consider it as an unknown. 
In order to determine all the unknowns,
at least 5 modes in the m-lines spectrum have to be identified. In general, {\it ie}  when
the thicknesses of the  layer 1 and layer 2 and the differences between the refractive 
indices of the different layers are large enough,  a spectrum contains
 $M>5$  modes.

\begin{figure}[htbp]
\centerline{
\includegraphics[width=7cm]{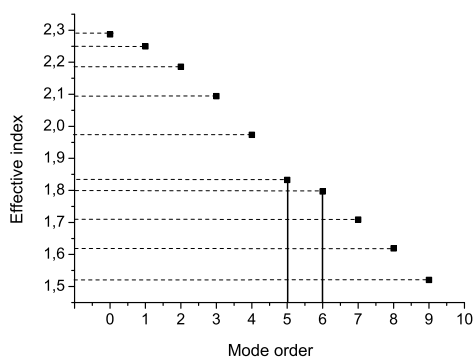}}
\caption{Effective indices of the guided modes of a two layer film defined by $n_1=1.9$, $d_1=1\ \mu$m, $n_2=2.2$ and $d_2=1.4\ \mu$m.}
\label{fig3}
\end{figure}

One difficulty consists in associating the correct equation to each mode.
As shown by equations~\ref{eq5} and \ref{eq6}, the effective index
of a mode depends on the type of guiding, either in one or in two layers.Hence
 a rupture in the spectrum  corresponding to the transition
 between these two types has to be determined.
 In the following we will
call $m_2$ the order of the first guided mode in two layers. Given that
the dispersion equations are transcendental, it is not possible
to derive a general analytical expression for $m_2$.  
 Numerical simulations with noisy datas, however,  showed  that   $m_2$  is the value of $m$ such that
$|N_m-N_{m-1}|>|N_{m+1}-N_{m}|$ (see figure~\ref{fig3}).  This criterion 
results in a correct value of $m_2$ 
in the case of two thirds of the simulated waveguides. For almost all the other cases,
the mismatch is equal to +1. Only one hundred of simulations from more than two millions lead to different
results. Then, in our data analysis protocol, we first
assume that $m_2$ obeys the criterion stated above and try then to solve the dispersion equations. If the program
does not converge, we substract one from the value of $m_2$ given
by the criterion and again  try  to solve the equations.
If this second step also fails, we may vary $m_2$
from 0 to $M$. It is  observed  that the program  only converges in the case of the real value of $m_2$. 

In order to retrieve the thin films chatacteristics, we  consider the
4 unknowns $n_1$, $d_1$, $n_2$ and $d_2$ and we
varie $m_1$. The   
$C_M^4$ systems of 4 dispersion equations are solved with a Newton-Raphson algorithm. 
For each value of $m_1$, we obtain $C\leq C_M^4$ sets of solutions.
  The 
synchronous angles ($\phi_{\rm calc}$) corresponding to each solutions 
are calculated with a bisection algorithm and we compute
the mean difference between these angles and the measured synchronous angles ($\phi_{\rm meas}$)~:
\begin{equation}
\label{eqsigma}
\sigma(m_1)=\sqrt\frac{\ds \sum_{i=1}^{C}\sum_{j=0}^{M-1}(\phi_{ij}^{\rm calc}-\phi_{j}^{\rm meas})^2}{M C^2}
\end{equation}
The minimum of $\sigma(m_1)$ gives the correct indexation. Finally, the solutions
are the mean values of the  solutions corresponding to the right indexation.

\subsection{Numerical tests}

\begin{figure}
   \begin{minipage}[c]{.49\linewidth}
 \centering      
\subfigure[Refractive index]{      
      \includegraphics[width=7cm]{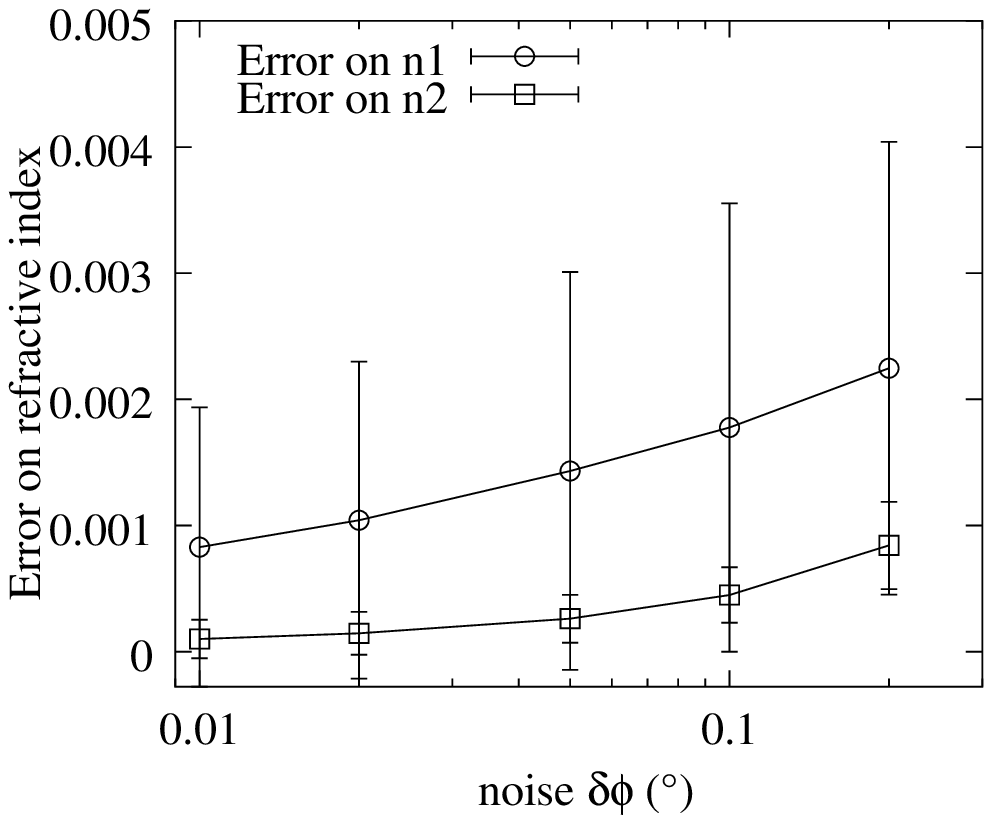}}☺
   \end{minipage} \hfill
   \begin{minipage}[c]{.49\linewidth}
 \centering
 \subfigure[Thickness]{     
      \includegraphics[width=7cm]{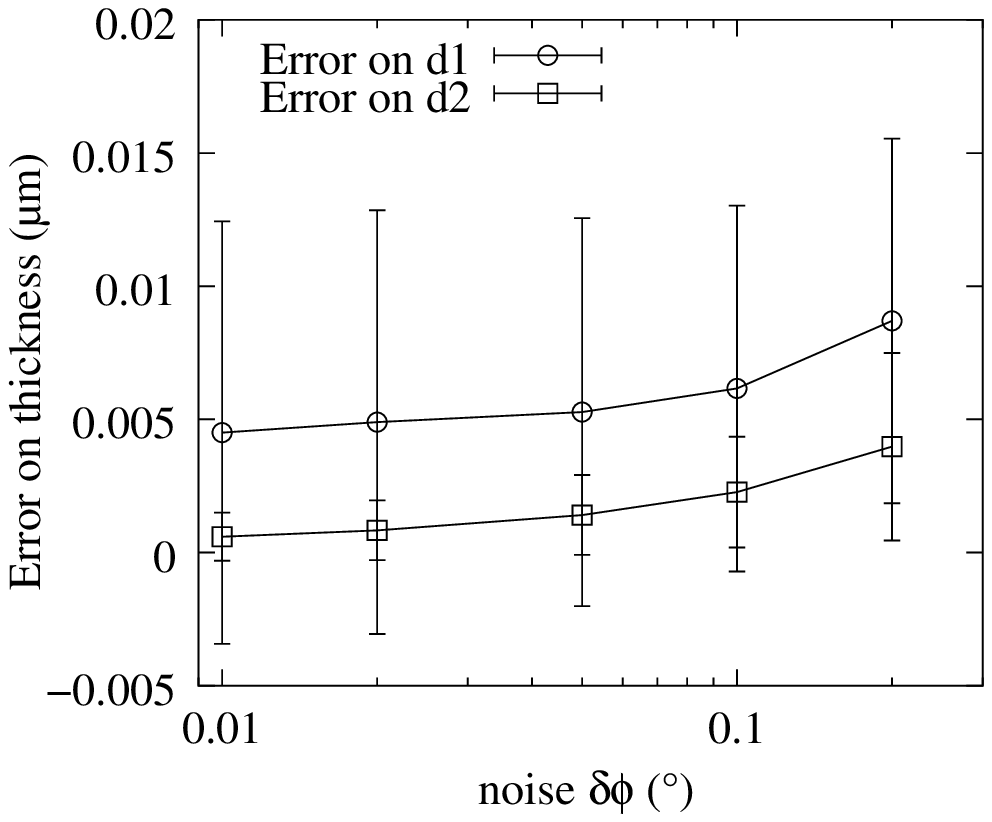}}
   \end{minipage}
   \caption{Mean errors on the refractive index and the film thickness as a function of noise.}
   \label{fig6}
\end{figure}

In order to estimate the accuracy of the numerical procedure described in the previous section,  10000 guides were simulated, $n_1$ varying  from 1.8 to 1.98 and $n_2$ from 2.12 to 2.30 by steps of 0.02, $d_1$ varying from 0.6~$\mu$m to 1.05~$\mu$m and $d_2$ from 0.8~$\mu$m to 1.25~$\mu$m by steps of 50~nm. These ranges correspond to the values of the layers which we process. For each waveguide $\{n_1,d_1,n_2,d_2\}$,  the set of synchronous
angles $\{\phi^j_{\rm th}\}$ is calculated and  a  noise 
randomly choosen in the range $[-\delta\varphi;\delta\varphi]$ is added in order to  obtain a  set of noisy synchronous angles $\{\phi^j_{\rm noisy}\}$. This set is used as input data of the numerical procedure. In order to get a statistical estimate of the accuracy, 100 sets of noisy angles were studied for each guide. The influence of noise was evaluated by varying $\delta\varphi$ between 0.01$^\circ$ and
0.2$^\circ$.
The distributions of errors on the different parameters for one waveguide follow a normal law. So, for each waveguide we consider that the error on one parameter is the mean value of the errors over the 100 noisy sets.
 Highest errors arise for guides with the smallest  index differences between the two layers and the smallest thicknesses. The effect of noise is shown in figure~\ref{fig6}. 
The errors on the refractive indices are plotted
in figure~\ref{fig6}a and the errors on the thicknesses in figure~\ref{fig6}b. The lengths of the errorbars is twice the standart deviation.   
The error on $n_2$ is smaller than the error on $n_1$
and does not exceed $1.10^{-3}$ in the considered noise range,
whereas the error on $n_1$ is of the order of a few $10^{-3}$.
The thicknesses are expected to be determined with a precision of the
order of 10~nm. 

\section{Characterization of PZT deposited on ZnO}

\subsection{Optical properties of the ZnO  layer}
Conducting transparent layers of Al doped ZnO  in hexagonal 
phase were deposited by rf magnetron sputtering at room temperature. 
  In our case, a
heat treatment  
of 650$^\circ$C is required for PZT  cristallization in  the perovskite phase.  In order to evaluate the possible effects
of this treatment,  a preliminary study on the ZnO  layers deposited on 25$\times$25~mm Corning 1737F glass
substrates was performed.

The zinc oxyde layers were grown by rf magnetron sputtering from a $\phi$3'' 
ZnO/Al$_2$O$_3$ (98/2 wt. \%) ceramic target. Prior to the deposition, a base pressure
lower than $5.10^{-7}$~mbar is reached and pure Ar is used as a  sputter gas at a chamber
pressure of $2.10^{-3}$~mbar during the deposition process. The applied rf power of 200~W results in  a growth
rate of approximatively 100~nm/min 
on axis at a target-substrate distance of 7.5~cm during 10~min.
As four samples were grown at the same time, 
the substrates were shifted from the center of symetry of sputtering chamber, 
and the thickness of ZnO decreased along the diagonale of the substrate. 
The ZnO target contains 2 wt\% of Al2O3, which corresponds to a [Al]/[Zn+Al] atomic 
ratio of about 3.3\%. In order to investigate the aluminium content within the thin films, 
an electron dispersion spectroscopy (EDS, JEOL 5800 LV equipped with a X-ray detector)
 mapping was performed on thin films grown 
on a silicon wafer (aluminium/zinc-free substrate) with the same procedure than that used 
in the present work. The analysis showed that  the aluminium content is homogeneous over 
the whole film surface  and its value is 3.3\%$\pm$0.5\%.
\begin{figure}
\begin{minipage}[c]{.49\linewidth}
\centering      
\subfigure[Refractive index]{
\includegraphics[width=7cm]{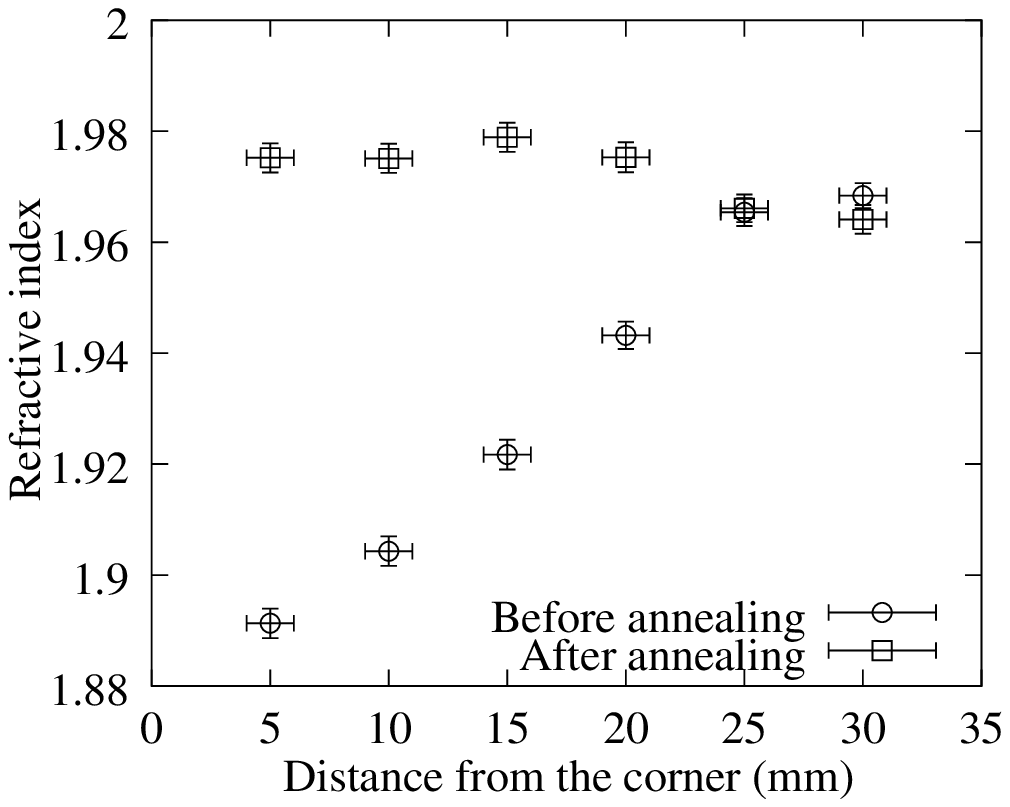}}
\end{minipage} \hfill
\begin{minipage}[c]{.49\linewidth}
\centering      
\subfigure[Thickness]{
\includegraphics[width=7cm]{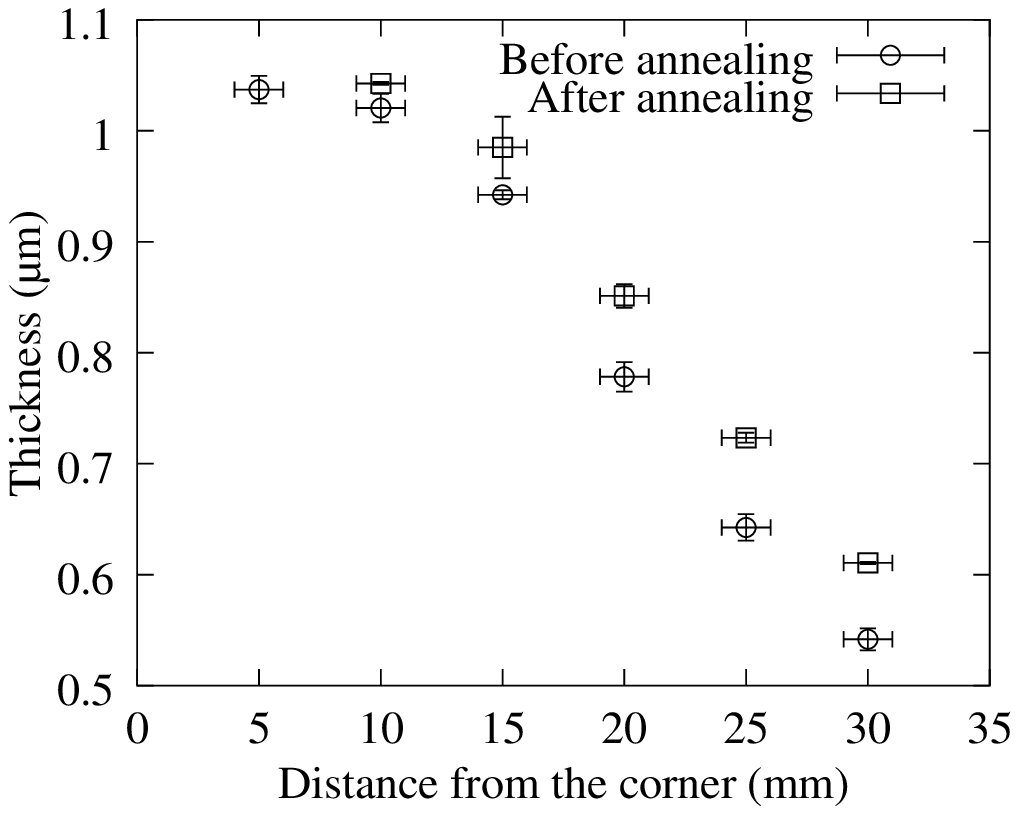}}
\end{minipage}
\caption{Evolution of  the refractive index and the film thickness along the diagonale of sample~1 before and after thermal annealing.}
\label{fig1}
\end{figure}

The films were characterized with m-lines spectroscopy using 
a set-up and data analysis program described elsewhere~\cite{cardin2005}.
The samples were  measured as deposited at
several points every 5~mm along the diagonale of the substrate. The evolutions of refractive index
and thickness of  sample 1 along this line are represented in the figure~\ref{fig1}a and \ref{fig1}b 
(curves with circles). These curves reveal the inhomogeneity of the film due to the deposition technique.
From one corner
to another, the thickness decreases from 1.04~$\mu$m
to 0.54~$\mu$m. Moreover it can be seen that the refractive index increases from 1.89 to 1.97.
  This implies that the cristalline structure of the ZnO 
is not the same across the film. The other samples show the same behavior, but sometimes,
the measurements at the lower
 thicknesses were difficult because of the lack of modes when the
thickness becomes too small. 

\begin{figure}[htbp]
\centerline{
\includegraphics[width=7cm]{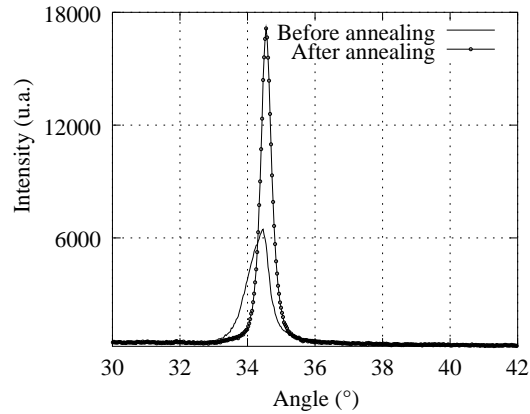}}
\caption{X-ray diffraction diagramm of the  ZnO film (sample~1) before and after annealing.}
\label{fig34}
\end{figure}

After this first characterization, the films were annealed at 650$^\circ$C during 3~min
and  cooled slowly in the oven during 3 hours. 
The results of the index and thickness of sample 1 are also
shown in the figures~\ref{fig1}a and \ref{fig1}b (curves with squares).
Obviously, the rapid thermal annealing does not change significantly the thickness,
however,  it  homogenizes
the refractive index of the film. Indeed, after annealing, the index 
 ranges between 1.975$\pm$2.10$^{-3}$ and
1.964$\pm$1.10$^{-3}$.
Previous studies~\cite{sun1999,ohya1994,mehan2004,alasmar2005} pointed out that  the annealing
improves the crystallinity  of the ZnO films 
by promoting the formation of stoichiometric ZnO.
The increase of the intensity and the decrease of the FWHM of the diffraction peak (xyz) 
well illustrate   this phenomenon (see figure~\ref{fig34}). Hence, the modification of refractive index with annealing can be related to this   crystallinity improvement.
 
\subsection{Optical properties of two layers waveguides}

\begin{figure}[htbp]
\centerline{
\includegraphics[width=7cm]{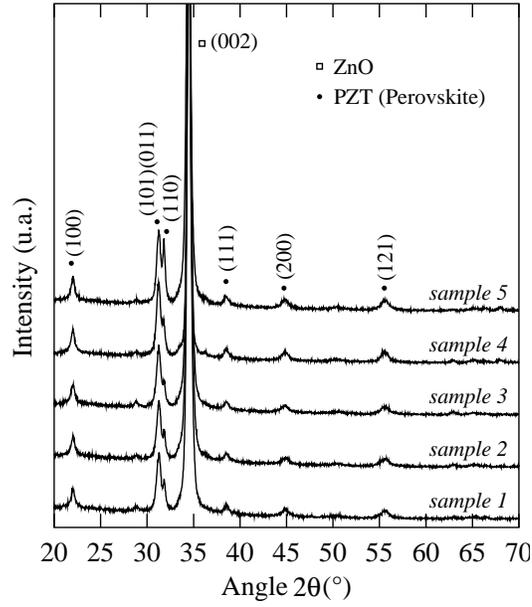}}
\caption{X-ray diffraction diagramm of the two layer films.}
\label{figxrd}
\end{figure}

\begin{figure}[htbp]
\centerline{
\includegraphics[width=7cm]{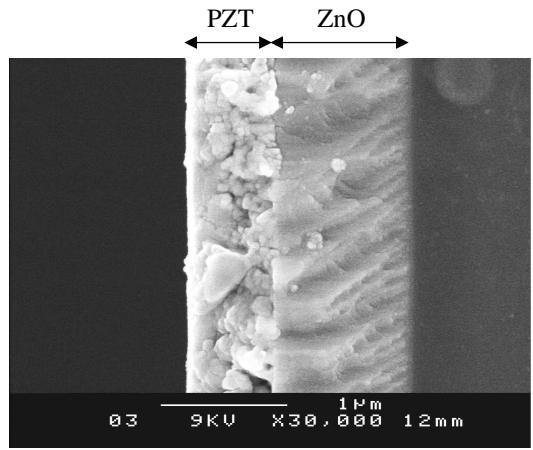}}
\caption{SEM photography of sample 3.}
\label{figmeb}
\end{figure}

PZT 36/64 thin films were elaborated by  
Chemical Solution Deposition  technique and were spin-coated  on the ZnO layer at 2000~rpm for samples 1 to 3 and  1000~rpm for samples 4 and 5.
A modified sol-gel process was used for the elaboration of
the precursor solution, which consisted of lead acetate dissolved in 
acetic acid, zirconium and titanium n-propoxide; 
ethylene glycol was added in order to prevent from crack formation during the annealing process. The deposited films were dried on a hot plate and a Rapid Thermal Annealing  procedure at 650$^\circ$C
resulting in the formation of a polycrystalline perovskite phase as shown by the XRD pattern (figure~\ref{figxrd}).
The samples were also studied with
scanning electron microscopy. An example is shown in figure~\ref{figmeb}.
The two layer structure appears clearly on this photography. The thickness
of the ZnO layer is approximatively 1.1~$\mu$m  and the thickness of the 
PZT layer is close to 0.7~$\mu$m. The values are
in agreement with the elaboration process.
\begin{figure}[htbp]
\centerline{
\includegraphics[width=7cm]{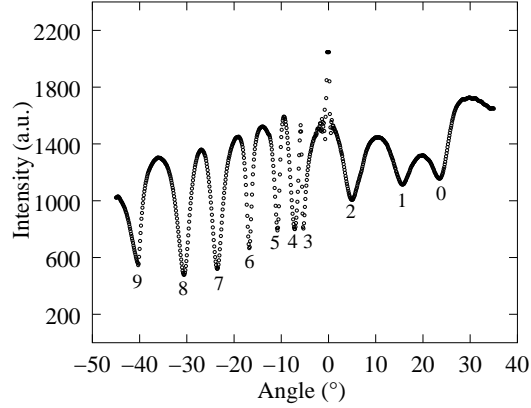}}
\caption{M-lines spectrum measured at one corner of sample 5.}
\label{fig7}
\end{figure}

The films were measured with m-lines at several points along the diagonale
of the sample. As an example, the spectrum of sample 5
is shown on figure~\ref{fig7}. The narrow peaks correspond to the waves guided in two layers
while the broad peaks correspond to waves guided in the PZT layer only. The broadening is not
a peculiarity of the two layer structure, it can be also observed with PZT single layers and
is probably due to the diffusion of light. The measurement given in  figure~\ref{fig7} was analysed with the previously
described procedure. Only the modes 4 to 8 were considered which 
can be located  with a precision of $\pm 0.05^\circ$, while the uncertainty on the position of the first
modes is of the order of 0.5$^\circ$ because of the broadening. 
From this measurement we  obtain~:
$n_1=1.977\pm 9.10^{-3}$ and $d_1=1.11\pm 4.10^{-2}\ \mu m$ for the ZnO layer and $n_2=2.36\pm 4.10^{-2}$ and $d_2=1.01\pm 7.10^{-2}$ for
the PZT. 
In order to verify the exactness of the results,  
 the values may be injected in the dispersion equations~\ref{eq5} and \ref{eq6}. Synchronous
angles in very good agreement with the measured angles are found (see Table~\ref{tab1}).
\begin{table}
\caption{Comparison of the measured and the calculated synchronous angles.} 
\begin{tabular}{l*{9}{c}}
modes  &1 & 2& 3 & 4 & 5 & 6  & 7 & 8 & 9  \\
\hline
measured ($^\circ$) &   23.5 & 15.60 & 4.90 & -5.30 & -7.20 & -10.80  & -16.70 & -23.40 & -30.40 \\
calculated ($^\circ$)  & 23.8 & 15.58 &  5.07 & -5.28 & -7.19 & -10.75 & -16.87 & -23.54 & -30.52  \\
\end{tabular}
\label{tab1}
\end{table}
 The differences never exceed 0.3$^\circ$, 
 corresponding to a maximum difference between measured and calculated effective indices of the order of 10$^{-3}$.

\begin{figure}
   \begin{minipage}[c]{.49\linewidth}
 \centering      
\subfigure[Refractive index]{
\includegraphics[width=7cm]{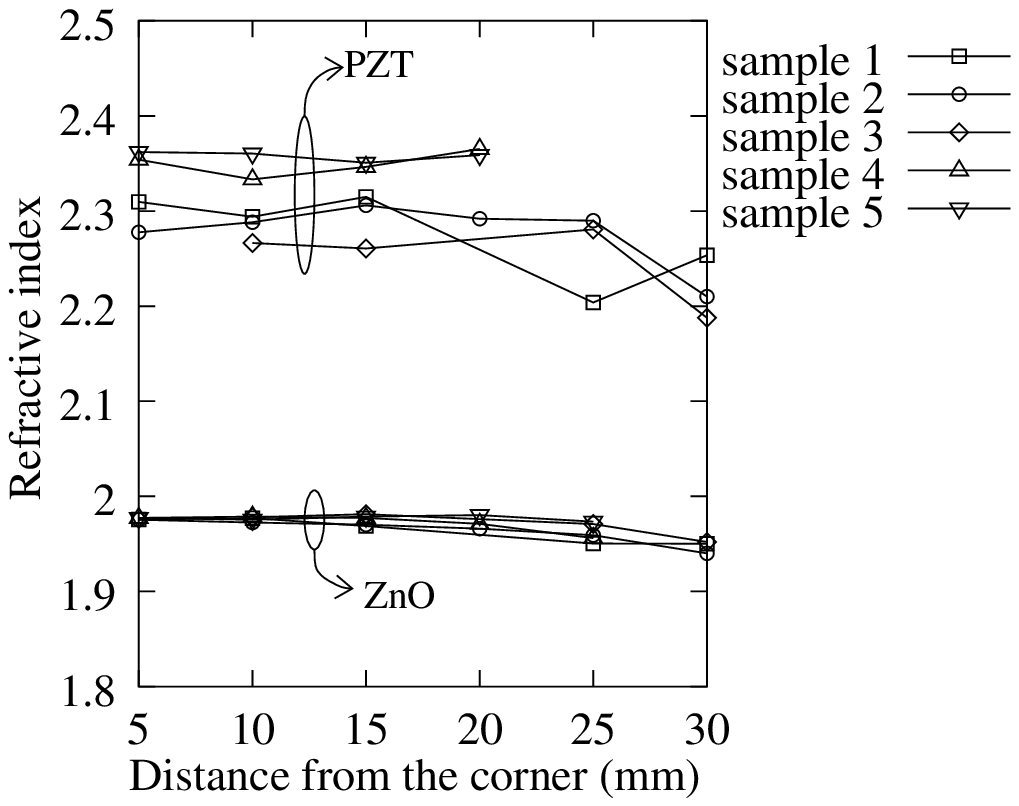}}
   \end{minipage} \hfill
   \begin{minipage}[c]{.49\linewidth}
 \centering
 \subfigure[Thickness of the ZnO layer]{    
      \includegraphics[width=7cm]{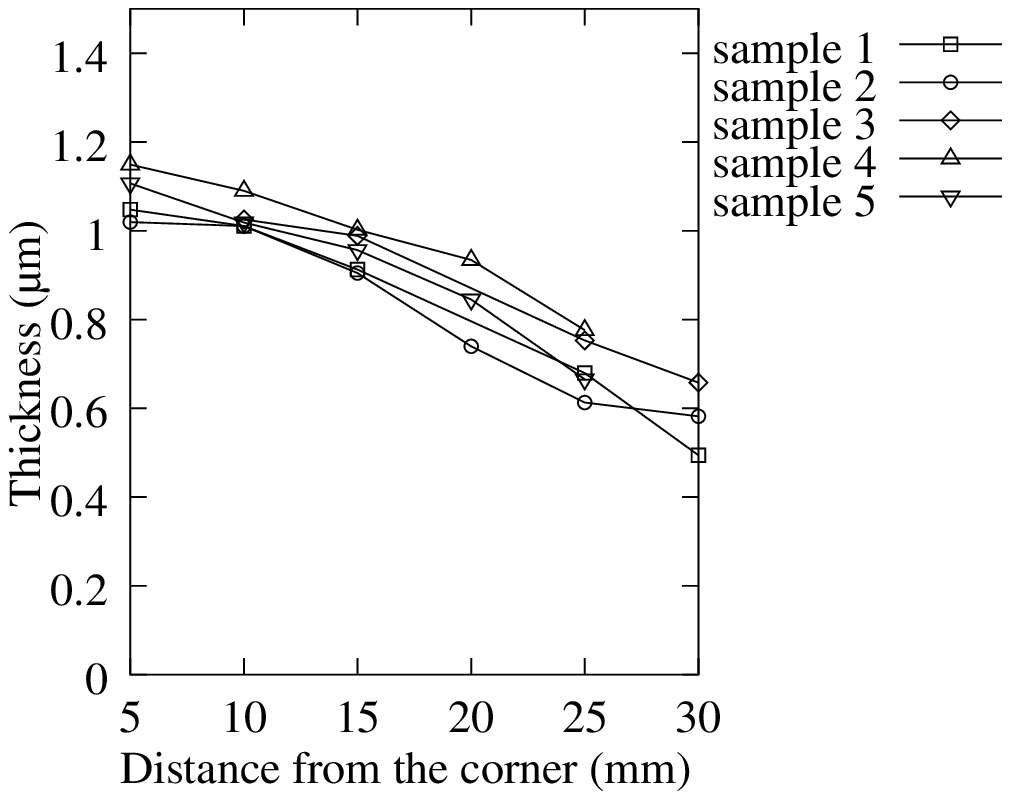}}
   \end{minipage}\\
 \centering
 \subfigure[Thickness of the PZT layer]{    
      \includegraphics[width=7cm]{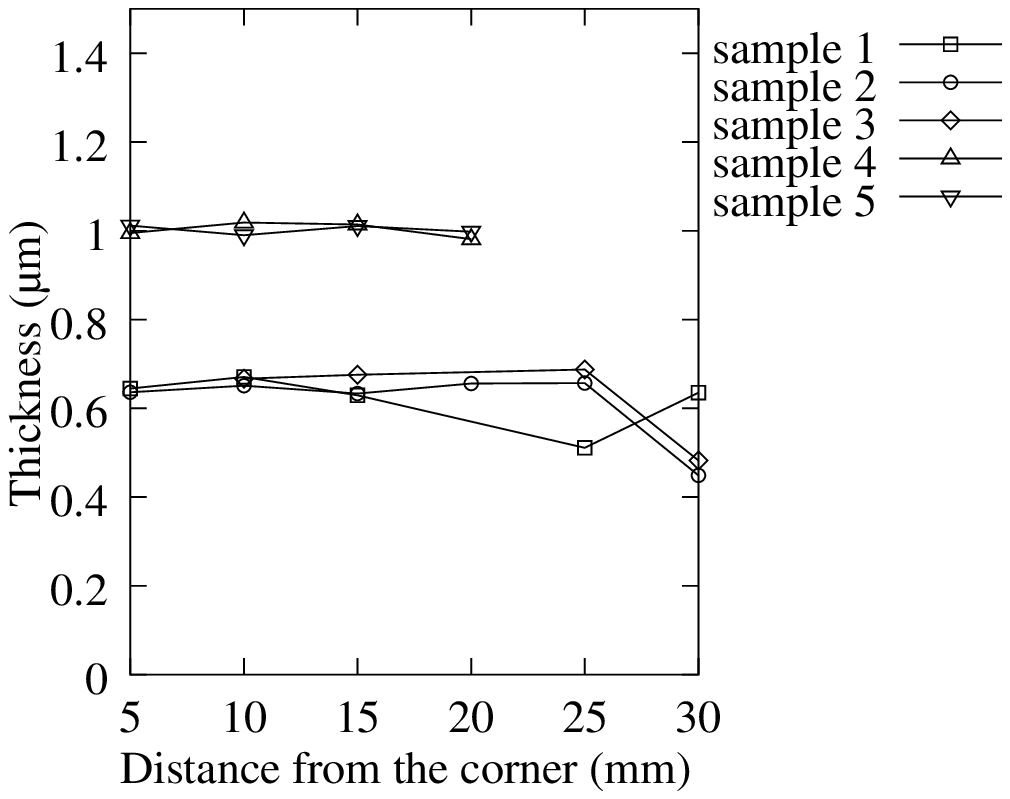}}
   \caption{Evolution of  the refractive index and the film thickness along the diagonale of the two layer samples.}
   \label{fig8}
\end{figure}

The results obtained with other samples are summarized in the figure~\ref{fig8}.
It was not possible to analyze each measurement, especially those corresponding to points where the ZnO layer  was too thin.
Everywhere else  a good agreement between the refractive indices
and thicknesses of the ZnO determined from the  single and the double layers was obtained. The variation is of
the order of 10$^{-2}$ for the index and 100~nm for the thickness in the worst case but in general remains lower than the uncertainties.  

\begin{figure}[htbp]
\centerline{
\includegraphics[width=7cm]{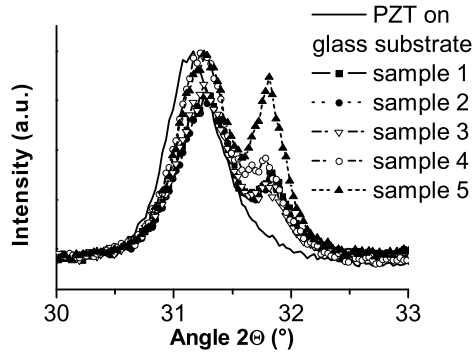}}
\caption{X ray diffraction diagramms}
\label{fig11}
\end{figure}

The measurement of the PZT refractive index reveals the influence resulting 
from the ZnO interface layer. A PZT film of the identical precursor composition, directly spin-coated on the glass substrate, has a refractive index close to 2.23, whereas the refracting index of the PZT deposited on ZnO is higher as can be seen in the figure~\ref{fig8}a.
 Moreover, the index varies from close to 2.36 for the thicker PZT films (deposited at 1000 rpm)
  to close to 2.30 for the thinner films (deposited at 2000 rpm). This indicates that the structure 
  of the film might be different. In order to verify this assumption, we compare in figure~\ref{fig11}, XRD measurements for the two cases. 
The unique peak at 2$\theta$=31$^\circ$ for PZT deposited on glass, corresponding to a rhomboedric structure is doubled when spin-coating the same composition PZT 36/64 on ZnO, indicating the appearance of a second phase corresponding to the tetragonal PZT structure. 
 
\section{Conclusions}
 
A two-layer PZT/ZnO wave-guide structure has been elaborated 
by rf magnetron sputtering and chemical solution deposition technique. 
The numerical tools for analyzing the m-lines spectra obtained from this 
two-layer system were developed and their efficiency in the presence of noise
 was demonstrated. In the case of a single ZnO layer, the benefice of a heat treatment 
 at higher temperatures in terms of homogenization of the refractive index has been shown. 
 The study of the PZT thin films revealed the sensitivity of the m-lines characterization method. 
 The different crystallization behavior of the ferroelectric resulting from the different structural 
 properties of the underlying layers (glass or ZnO), could be observed by m-lines spectroscopy as a 
 change of the PZT refractive index. This appears to be very important for the design
 of single mode waveguides since the thickness of the guiding layer 
 is related to the difference between the  refractive indices of the guiding and the confining
 layers. The presented work can be considered as a first step
 towards the characterization of three layer composite structures by m-lines spectroscopy, 
 also resulting in the possibility to determine the refractive index
 as a function of an applied electric field and to deduce
  the electro-optical coefficient. 



\end{document}